# Voltage Security Analysis of VSC-HVDC Transmission Lines

Chen WANG[1, 2], Chetan MISHRA[2]
[1] Virginia Polytechnic Institute and State University
[2] Dominion Energy
USA

**SUMMARY**

Due to generation retirements and growth of renewable energy sources integration, there will be widespread changes in the real and reactive power flow in nowadays power systems. These changes also bring about challenges on system operation and stability. High voltage direct current (HVDC) technology seems to be a promising solution to these new challenges. A study on the impact of the replacement of 500KV AC transmission lines by VSC-HVDC transmission line on system voltage security is conducted. The analysis is based on reactive load margin method and implemented on one of the Dominion Energy planning models. Different control schemes of HVDC are considered. Using k-means clustering method, three representative zones within the network are selected. The results corresponding to them are demonstrated and discussed. It is shown that HVDC lines with P-V control remarkably improve system voltage security, while those with P-PF control scheme have negative effects.

chenwang@vt.edu

# 1. INTRODUCTION

Nowadays, power systems work closer to their operating limits and stability boundaries due to efficiency and economic reasons. At the same time, more and more renewable energy sources have been integrated into both the transmission and distribution systems[1]. All these factors increase the danger of system voltage collapse and triggering large blackouts[2][3]. Therefore, system voltage stability has become one of the major concerns for power system planning and operation. Voltage source converter (VSC) based HVDC, given its flexibility of power flow and voltage controls, can be a suitable solution[4]. With it implemented in the system, there can be an improvement in system's capability to keep bus voltages within acceptable limits during normal operation as well as when subjected to faults.

Considerable research efforts have been devoted to VSC-HVDC control schemes and their impacts on system voltage security. In [4], an appropriate control scheme of VSC-HVDC is proposed. The positive impact of its application on long-term system voltage stability is assessed and validated. Researchers in [5] guarantees sufficient stability margin of the HVDC integrated system by using sensitivity between VSC control input and system voltage stability margin to guide the control algorithm. In [6], a comparison is conducted between a new ac transmission line and a new VSC-HVDC line. The results show that the voltage support capability of VSC-HVDC can help the system from losing synchronism caused by voltage collapse. Using P-V curve, Study in [7] demonstrates that loss of stability margin can be minimized by proper control of VSCs when currents reach their limits.

In this paper, we are also considering use reactive load margin as the metric to system voltage security, which is widely utilized nowadays. In order to demonstrate VSC-HVDC lines' impact on the overall system, the network in Dominion Energy territory is used as the study case. This system is partitioned into several zones. The average of buses' reactive load margins in each zone is evaluated. The results of cases with different VSC-HVDC lines are clustered using k-means algorithm. Three representative cases are selected from these clusters and analyzed.

The remainder of the paper is organized as follows: Section II describes the voltage stability analysis and k-means method. In section III, study cases are described. Simulation and computation results are presented in section IV. Finally, section V concludes the paper.

# 2. METHODOLOGY
## A. Reactive Load Margin

Magnitude of voltages is intuitively used to illustrate system voltage security[8]. However, it does not necessarily give the correct picture of voltage health in a network. In an overly compensated system, even a 0.9 p.u. voltage on a bus could mean a danger of collapse. This can be seen from the scatterplot below between the max Q that can be extracted at each bus (negative for extraction) and the voltage at the point of collapse for all the buses in Dominion in a summer peak planning model. Here it can be seen that some buses collapse at even 1 p.u.

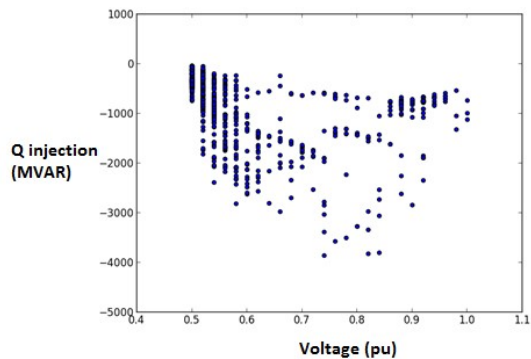

*Figure 1. Typical Q-V Scatterplot in Dominion Peak Summer*

Voltage security refers to how far the operating condition is from the point of collapse or loss of equilibrium for a given change in load[9]. The buses in a power system are connected by transmission lines with each line's impedance [10] limiting the maximum allowable power flow over it. A system has multiple voltage control devices like generator AVRs, SVCs, etc which dynamically change their reactive power injection to the system in order to maintain voltage at a pre-defined bus through control action. The electrical distance (depends on impedance) of a bus from such reactive power resource determines its accessibility to that resource. An increasing reactive load on a particular bus will be met by some and not necessarily all such resources in the system. The collection of such resources that participate to match the increasing reactive power demand at a particular bus forms the reactive reserve basin (RRB) for that bus. The size of the RRB reduces as kV level increases or electrical distance to those resources increases. Each such resource has a limit on how much reactive power it can provide. Thus, on occasions when somehow the reactive reserve on the RRB for some region is exhausted, the voltage becomes highly insecure.

Q-V analysis can be done at each bus in the network to understand how well controlled the voltage is at each bus and up to what extent[11]. The analysis involves adding a fictitious generator on the study bus and then reducing its voltage set point which the generator achieves by absorbing reactive power from the network. The Q value at which the voltage collapses gives a sense of the security margin at that bus. Also, the number of reactive resources that where loss of voltage control happens leading up to the voltage collapse at the study bus gives an idea about the size of the RRB. A typical QV curve is shown below[12]. Larger the Q margin, better the voltage health is at that bus.

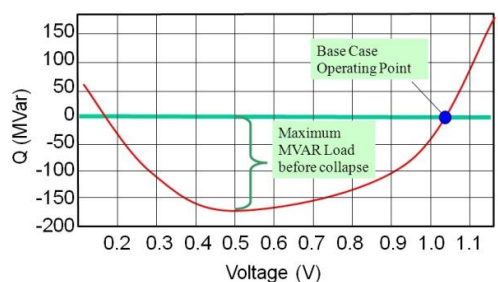

*Figure 2. Typical QV Curve*

**B. k-means Clustering**

K-means is a classical clustering algorithm used for unsupervised machine learning[13]. The basic idea is to iteratively assign data points into groups so that their 'distances' to their own fictitious group centers are minimized in general.

In this paper, the k-means method is used to find groups of HVDC line cases that have similar impacts on reactive load margins of the zones in the system. The features for each HVDC line case are the differences of average Q-margin between HVDC case and base case of each zone. The metric is the Euclidian distance for simplicity reason and limited dimension of the features.

**3. STUDY CASE**

The simulation and computation are based on PSS/E and Python, respectively. One planning model of Dominion system is used as the base case. The power flow model of VSC-HVDC in PSS\E is constructed, which is comprised of one voltage source converter (VSC) on each end connected by a DC line. For each 500 kV AC line upgrade to HVDC, two new cases are created with HVDC operating at P-PF and P-V control scheme respectively (total 90 cases). The analysis on the original base case as well as each of these created cases would be done and compared.

## 4. RESULTS

The clustered Q-margin difference between P-PF control VSC-HVDC and base case is shown in the following figure. There are five clusters in total and each of them is differentiated from the others with black vertical lines in the figure. The colors indicate the percentage of the Q-margin increases from the base case.

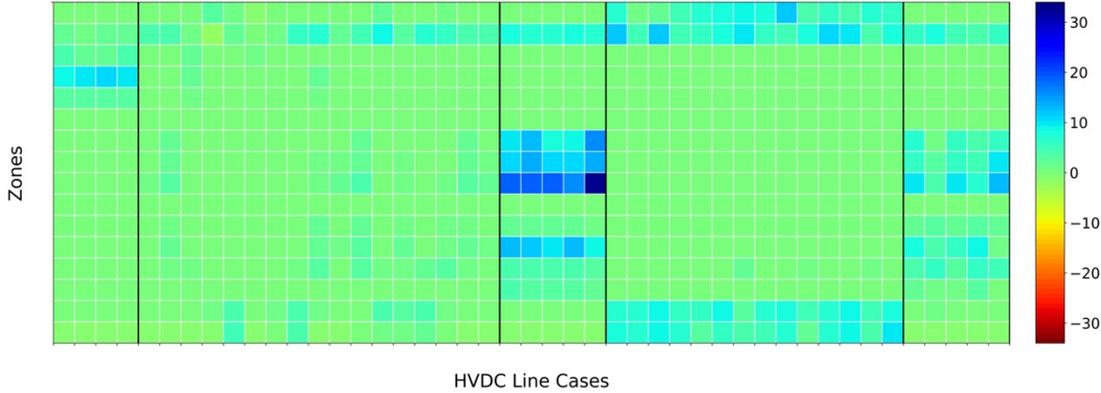

*Figure 3. Q-margin percentage difference between AC case and P-V control HVDC case*

The clustered Q-margin difference between P-PF control VSC-HVDC and base case is shown in the following figure.

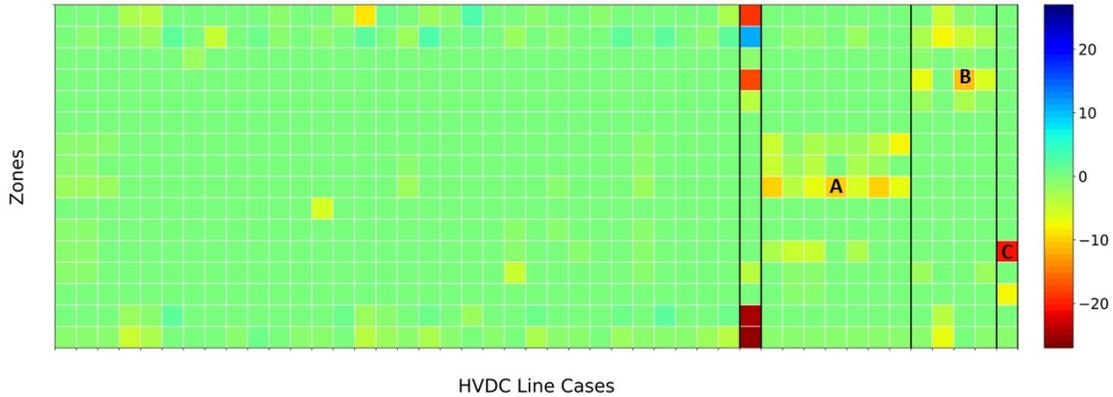

*Figure 4. Q-margin percentage difference between AC case and P-PF control HVDC case*

Generally, the VSC-HVDC lines with P-V control can help the system to reach larger Q-margin. The largest increase can even be 30%, which is a great improvement on system voltage security. At the same time, those with P-PF control cannot provide this impact. The Q-margin mostly stay the same as base case, if not decreasing.

Three cases (labeled as A, B, and C in Figure 4) with corresponding zones are selected for further analysis.

### Case A:

In this case, zone 13 used to suffer from voltage issues in the past. After installation of STACOMS, the issues are mitigated. It is fed by three 500KV AC lines in reality. Within the base case we used, this zone absorbs real power from zone 12 and provides zone 12 and zone 15 with reactive power support in normal operation condition. The HVDC line is assumed to be replacing one of the AC lines connecting this zone and zone 12.

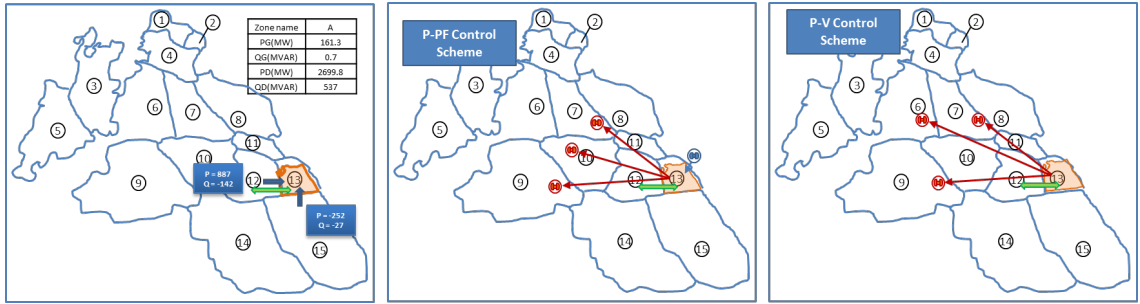

*Figure 5. Example case: Case A*

The middle figure shows the effect on generator reactive power output made by VSC-HVDC under P-PF control. From the figure we can see that with the P and Q fixed on the HVDC line, generators distributed in three zones cannot reach maximum reactive power output as reactive loads in zone 13 increases. So these loads have to be satisfied by local generators. This is the reason for Q-margin decrease. (The red generator symbol means that these generators used to output maximum reactive power in base case but can no longer provide that in HVDC case; the blue generator symbol means the opposite.)

The right figure indicates that VSC-HVDC with P-V control can work as STATCOM to support the reactive loads locally in the zone. Even that the remote reactive power sources are decoupled, the Q-margin still increases.

**Case B:**

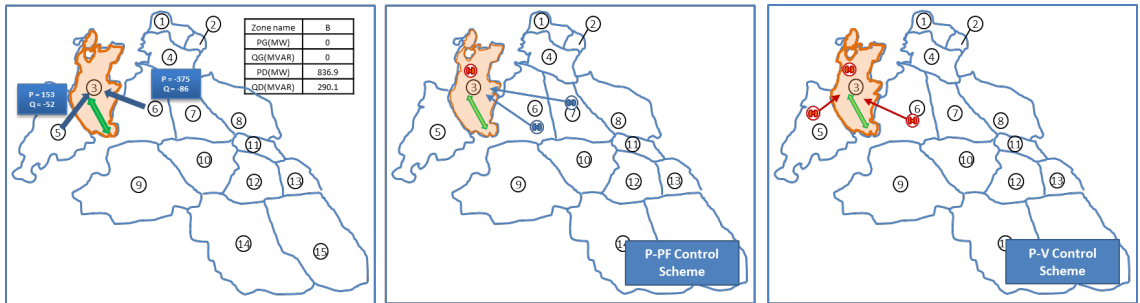

*Figure 6. Example case: Case B*

In this case, zone 3 is supported by large generation locally. However, there is also large economic land resource available for renewable energy sources development, which could cause voltage fluctuations requiring better voltage controllability in the future.

Considering P-PF control, voltage support from the far side of the HVDC line is significantly weakened. The Q-margin decrease severely, which could cause poorer voltage controllability. With P-V control, the lost controllability is compensated by HVDC link. This results in an overall marginal improvement in Q margin.

**Case C:**

In this case, there have already been voltage issues in zone 11 because of generation retirement. Putting VSC-HVDC with P-PF control is equivalent to removing the reactive power support from zones 9, 13, and 14, which causes major reductions in Q-margin. Even though more generators' reactive power output is attracted from zone 8 and 10, due to inner network transmission capacity, the reactive power cannot reach the bus with increased reactive load. For P-V control, on the other hand, the HVDC link itself can serve as STATCOMs. Even though there is reduce outside reactive power resource, the overall Q-margin is increased.

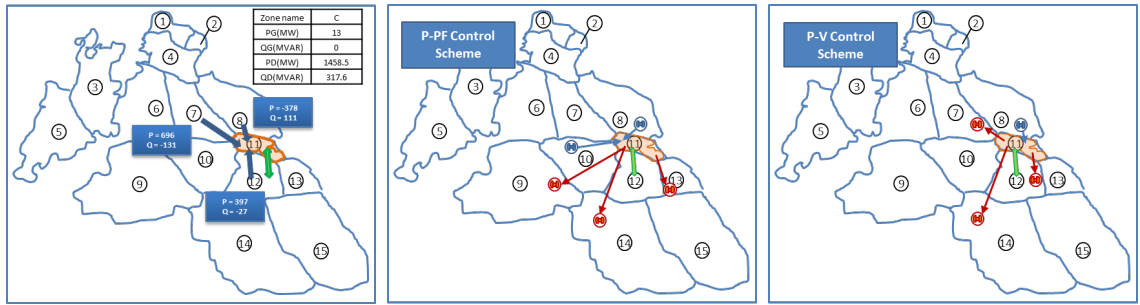

*Figure 7. Example case: Case C*

## 5. CONCLUSION

In this paper, the voltage security of Dominion system with selected AC lines upgraded to VSC-HVDC links is analyzed with reactive load margin. The simulation is run on PSS/E and the results are discussed. To sum up, the impact of replacing an AC line with an HVDC on voltage security can be understood in the following way. Take an example of a load area in the system connected to the rest of the system through two AC lines. One AC line is a 500 kV low impedance line connecting it to a large pool of generation. Another line is a high impedance 230 kV line connecting it to a small local generation. In this scenario, the 500 kV line helps this area stay secure by bringing it closer to a big reactive basin. Now, upgrading this line from AC to DC operating at P-PF fixes the reactive power flow over this 500 kV corridor. This can be thought of as removing the AC line and replacing it with fixed injections on both ends which effectively increases the electrical distance of the load area from the large pool of generation it earlier had access to thus significantly reducing the size of its RRB and therefore negatively impacts the voltage security of this area. Let us now take the example of changing the control scheme of the HVDC to P-V. In this scenario, the HVDC tries to maintain the voltage on its endpoint lying in the load area. This can be thought of as removing the AC line and replacing it with real power load with STATCOMs on both ends. Now, depending on the reactive power limit of the VSC on the load area side as compared to the size of the RRB and/or maximum Q that could be extracted over the existing 500 kV AC line, the voltage security may be improved or decreased.